\documentclass[sigconf,nonacm]{acmart}

\usepackage{pgfplots}
\usepackage{algorithm}
\usepackage{algpseudocode}
\pgfplotsset{compat=1.18}
\usetikzlibrary{calc}
\renewcommand\footnotetextcopyrightpermission[1]{}

\begin{document}

\title{Balancing Privacy and Efficiency: Music Information Retrieval via Additive Homomorphic Encryption}

\author{William Zerong Wang}
\authornote{This work was conducted while the author was a research intern at the University of Washington.}
\affiliation{%
  \institution{Columbia University}
  \country{USA}
}
\email{william.z.wang@columbia.edu}

\author{Dongfang Zhao}
\affiliation{%
  \institution{University of Washington}
  \country{Tacoma, USA}
}
\email{dzhao@washington.edu}

\renewcommand{\shortauthors}{Wang and Zhao}

\begin{abstract}
Modern music retrieval runs on vector embeddings, and once these embeddings are shared for search or matching they can be copied, probed, or used to train generative models. Fully homomorphic encryption can compute on them but is impractical at scale because of ciphertext--ciphertext multiplication and bootstrapping. We observe that when only one operand is encrypted, the query or the database, similarity search reduces to ciphertext addition and ciphertext--plaintext multiplication, which additive homomorphic encryption supports cheaply (through Paillier, or CKKS restricted to additive operations). Building on this observation, we (i) implement two music-specific inference attacks and quantify the privacy--utility tradeoff of provable mitigations, (ii) introduce structure-aware additive primitives with learned per-block weighting at no extra cryptographic cost, and (iii) show across four audio datasets that additive search preserves nearest-neighbor rankings exactly while avoiding the ciphertext--ciphertext multiplication and bootstrapping of full-depth FHE and scaling far better in embedding dimension than a true-additive Paillier baseline.
\end{abstract}
\maketitle

\section{Introduction}
\label{sec:intro}
Music data poses distinctive privacy challenges in the era of generative AI. Systems such as OpenAI's Jukebox~\cite{dhariwal2020jukebox}, Google's MusicLM~\cite{agostinelli2023musiclm}, MusicGen~\cite{musicgen2023}, and AudioLM~\cite{audiolm2023} generate high-fidelity audio conditioned on genre, instrumentation, or text, and they rely on learned vector embeddings for synthesis, classification, and recommendation~\cite{briot2017deep}. Those same embeddings introduce exposure: once released, they can be sampled, remixed, or reverse-engineered without any access to the original audio. Protecting these representations while keeping them useful is therefore central to secure and ethical machine learning on music.

Recent incidents show the stakes are concrete rather than hypothetical. In 2023 the viral track ``Heart on My Sleeve'' cloned the voices of Drake and The Weeknd and drew millions of views before removal at the request of Universal Music Group~\cite{coscarelli2023ai}. Commercial exposure is growing in parallel: in December 2025 Warner Music Group sued the retailer PacSun over unlicensed recordings in social media posts, with reported damages in the tens of millions~\cite{wmg_pacsun2025}. Unreleased catalogues are also repeatedly exfiltrated, from a ransomed archive of Radiohead recordings~\cite{radiohead_minidiscs2019} to leaked demos from Madonna's \emph{Rebel Heart}~\cite{madonna_rebelheart2015}.

These episodes share a lesson: a catalogue is a high-value, actively targeted asset, and every copy that exists in the clear is an added attack surface. Increasingly, the copy that circulates is not the audio but the embedding, which is shared for search and matching. Our aim is not to prevent audio piracy; it is to ensure that when a catalogue must be shared for computation, its embeddings need not be exposed as one more plaintext copy. Queries deserve protection too, since listening data reveals personal and emotional states.

Protecting the embeddings themselves therefore requires cryptography, since licensing and watermarking offer little protection for abstract numerical vectors. The operation to protect is vector similarity search, the fundamental primitive of music information retrieval (IR), which matches and ranks audio by content and underpins applications from audio identification~\cite{shazam2003,casey2008content} to recommendation over vector databases~\cite{milvus2021,singlestorev,solmaz_icict24}. Fully homomorphic encryption (FHE) has been studied for encrypted similarity search but remains costly at retrieval scale~\cite{cgentry_stoc09,cheon2017homomorphic}, because it relies on ciphertext--ciphertext multiplication and periodic bootstrapping~\cite{ckksboot2018}.

Our starting point is that this full generality is usually unnecessary, since multiplying two ciphertexts is not required. Prior work shows that only one operand, the stored database vectors or the query vector, needs encryption~\cite{prism_sigmod21,otawose_sigmod23,secureknn2009,elmehdwi2014}. Music IR fits this pattern: depending on the application either the query or the database must remain confidential, but rarely both at once. When only one operand is encrypted, the inner product $\langle \mathbf{x},\mathbf{y}\rangle=\sum_i x_i y_i$ is computed entirely from ciphertext additions and ciphertext--plaintext multiplications, so no multiplicative depth is consumed and bootstrapping is never triggered. We call any homomorphic scheme run under this restriction \emph{additive homomorphic encryption (AHE)}, realized by a genuinely additive scheme such as Paillier~\cite{paillier1999} or by CKKS restricted to additive operations~\cite{cheon2017homomorphic}.

The additive reduction is not itself new to cryptography; single-operand encryption underlies several privacy-preserving learning systems~\cite{prism_sigmod21,otawose_sigmod23,zhao2025note,ipfe2015}. Our contribution is to turn it into a practical, fully analyzed system for music IR:
\begin{itemize}
\item \emph{A new method.} An encrypted similarity search framework that runs entirely in the additive setting, supports both an encrypted-query and an encrypted-database deployment, and adds two structure-aware primitives, the Blocked Inner Product and the Weighted Hierarchical Inner Product, whose per-block weights are learned from training data and applied at no additional cryptographic cost.
\item \emph{Theoretical significance.} Proofs that the computing server learns nothing under standard chosen-plaintext (IND-CPA) security (Proposition~\ref{prop:server}), that blocking and weighting are exactly correct at a cost of at most $k$ scalar multiplications (Propositions~\ref{prop:blocked} and~\ref{prop:whip}), and that Gaussian noise on the score channel gives $(\varepsilon,\delta)$-differential privacy per query with a closed-form bound under adaptive composition (Proposition~\ref{prop:gm} and Corollary~\ref{cor:comp}); we also formalize two music-specific inference attacks that these guarantees are designed to contain.
\item \emph{Experimental results.} Across four audio datasets, encrypted additive search reproduces plaintext rankings exactly, runs up to $2\times$ faster than full-depth encrypted cosine and up to $3.7\times$ faster than Paillier at 1024 dimensions, improves normalized discounted cumulative gain (nDCG@10) by 0.12 to 0.27 with learned weights when blocks differ in relevance, and drives both attacks toward chance under calibrated noise at a quantified utility cost.
\end{itemize}

The remainder of this paper is organized as follows. Section~\ref{sec:background} reviews background and related work; Section~\ref{sec:method} develops the additive framework with full proofs; Section~\ref{sec:eval} reports the evaluation; and Section~\ref{sec:conclusion} concludes.

\section{Background and Related Work}
\label{sec:background}

\subsection{Homomorphic Encryption and Secure Computation}
\label{subsec:he}
Homomorphic encryption lets a party compute on ciphertexts so that decryption yields the correct result on the underlying plaintexts. Fully homomorphic encryption evaluates arbitrary circuits~\cite{cgentry_stoc09}, and modern schemes span integer arithmetic (BGV~\cite{bgv2012}, BFV~\cite{bfv2012}), Boolean logic (TFHE~\cite{tfhe2020}), and approximate real arithmetic (CKKS~\cite{cheon2017homomorphic}), with mature libraries such as Microsoft SEAL~\cite{sealcrypto}, Lattigo~\cite{lattigo}, OpenFHE~\cite{openfhe}, and TenSEAL~\cite{benaissa2021tenseal}. Two operations matter for our purposes: ciphertext addition, and ciphertext--plaintext multiplication by a public constant; both are inexpensive and add little noise. In contrast, ciphertext--ciphertext multiplication is expensive, consumes multiplicative depth, and eventually forces bootstrapping to refresh accumulated noise~\cite{ckksboot2018}. Paillier~\cite{paillier1999} is additively homomorphic by construction, exact over integers, but requires a modular exponentiation per scalar multiplication; CKKS supports approximate arithmetic over packed real vectors, which makes vectorized inner products efficient.

Homomorphic and multi-party techniques have been applied to neural inference (CryptoNets~\cite{gilad2016cryptonets}, Gazelle~\cite{juvekar2018gazelle}), to general privacy-preserving learning (SecureML~\cite{mohassel2017secureml}), to secure aggregation~\cite{bonawitz2017practical}, and to federated learning with additively homomorphic encryption~\cite{hardy2017private,zhang_atc20}, including on edge devices~\cite{mkhan_icws24}. Our work stays strictly additive, avoiding both ciphertext--ciphertext multiplication and bootstrapping, and uses full-depth CKKS only as a baseline.

\subsection{Encrypted Similarity Search and Secure kNN}
A long line of work computes over encrypted data, from searchable symmetric encryption~\cite{dsong_sp00} and encrypted databases such as CryptDB~\cite{popa2011cryptdb} and Arx~\cite{arx_vldb19}, to secure $k$-nearest neighbor (kNN) search on outsourced databases~\cite{secureknn2009,elmehdwi2014} and secure approximate nearest neighbor search~\cite{sanns2020}, inner-product functional encryption~\cite{ipfe2015}, and private information retrieval~\cite{pir1995}. Several systems show that encrypting a single operand suffices for efficient inner-product evaluation~\cite{prism_sigmod21,otawose_sigmod23,zhao2025note}. We build on this single-operand insight but target music embeddings specifically, add structure-aware primitives, and analyze the leakage that persists in the decrypted score channel.

\subsection{Approximate Nearest Neighbor Search and Vector Databases}
Large-scale retrieval relies on approximate nearest neighbor (ANN) indexes, including graph methods (HNSW~\cite{hnsw}), quantization (product quantization~\cite{pq2011}, ScaNN~\cite{scann2020}), and disk-resident indexes (DiskANN~\cite{diskann2019}), exposed through systems such as FAISS~\cite{johnson2017faiss}, Milvus~\cite{milvus2021}, Pinecone~\cite{pinecone}, SingleStore-V~\cite{singlestorev}, pgvector~\cite{pgvector}, Qdrant~\cite{qdrant}, and Weaviate~\cite{weaviate}. These systems rarely offer efficient privacy; our additive evaluation is index-agnostic.

\subsection{Music IR, Audio Embeddings, and the Requirements of Music Retrieval}
\label{subsec:musicir}
Music retrieval is built on learned audio embeddings, including general audio models (YAMNet~\cite{plakal2020yamnet}, VGGish~\cite{vggish2017}, PANNs~\cite{panns2020}, OpenL3~\cite{openl3_2019}), speech models (wav2vec~2.0~\cite{wav2vec2020}, HuBERT~\cite{hubert2021}), and music-specific and language-audio models (MERT~\cite{mert2024}, CLAP~\cite{wu2023clap}). Classical audio fingerprinting~\cite{shazam2003,chromaprint,panako2014} and neural fingerprinting~\cite{neuralfp2021} address exact recognition, whereas we protect the embeddings that power fuzzy retrieval.

Music retrieval is not image or text retrieval with new data. Music is temporal, so sequential structure matters and longer embeddings are needed~\cite{ren2020pirhdy}; it is layered, since orchestral works and remixes overlay many sources; and it demands invariance to transposition, version~\cite{yesiler2020accurate}, tempo, and quality, with tolerance for fuzziness in genre and mood~\cite{casey2008content}. Interactive retrieval also demands low latency: commercial recognition systems identify a track within seconds against millions of references~\cite{shazam2003}. High dimensionality and tight latency make naive encryption impractical; this layered, temporal structure also motivates the blocked and weighted primitives of Section~\ref{sec:method}.

\subsection{Differential Privacy}
\label{subsec:dp}
Differential privacy (DP) quantifies how much a single record can influence a randomized output. A mechanism $\mathcal{M}$ is $(\varepsilon,\delta)$-differentially private if, for all neighboring databases $D,D'$ differing in one record and all measurable output sets $S$, $\Pr[\mathcal{M}(D)\in S]\le e^{\varepsilon}\Pr[\mathcal{M}(D')\in S]+\delta$~\cite{dwork2006calibrating}. Smaller $\varepsilon$ means stronger privacy. The Gaussian mechanism achieves this by adding noise calibrated to the $\ell_2$-sensitivity of the released function~\cite{dwork2014algorithmic,balle2018improving}, and when a client issues many adaptive queries, advanced composition bounds the total privacy loss sublinearly in the number of queries~\cite{dwork2010composition}. Differentially private stochastic gradient descent (DP-SGD)~\cite{dpsgd2016} applies these tools at training time; we apply them to the \emph{score channel} of retrieval and measure the resulting tradeoff empirically.

\section{Similarity Search under Additive Homomorphic Encryption}
\label{sec:method}
This section formalizes the additive setting, the threat model and what the server can learn, the attacks on the score channel, the mitigations and their guarantees, and the structure-aware primitives (Sections~\ref{subsec:configs} to~\ref{subsec:whip}).

\subsection{Setting, Notation, and the Additive Reduction}
\label{subsec:configs}
Let $\mathbf{x}\in\mathbb{R}^d$ be a query embedding and $\{\mathbf{y}_j\}_{j=1}^{N}$ a database of embeddings, with similarity score $s_j=\langle\mathbf{x},\mathbf{y}_j\rangle$. All embeddings are $\ell_2$-normalized, so inner-product ranking coincides with cosine ranking and with Euclidean ranking on unit vectors.

\paragraph{Notation.}
For an integer $n$, $[n]$ denotes the index set $\{1,\dots,n\}$; $\langle\cdot,\cdot\rangle$ is the Euclidean inner product; $\lVert\cdot\rVert_2$ is the $\ell_2$ norm; and $I_N$ is the $N\times N$ identity matrix. $\mathsf{KeyGen}$ outputs a key pair $(pk,sk)$; encryption $Enc$ uses the public key $pk$, decryption $Dec$ uses the secret key $sk$, and $Enc(\mathbf{y})$ denotes the coordinate-wise (or packed) encryption of a vector $\mathbf{y}$. A scheme is IND-CPA secure (indistinguishable under chosen-plaintext attack) if no probabilistic polynomial-time adversary can distinguish encryptions of two messages of its choice with non-negligible advantage.

\begin{definition}[Additively homomorphic encryption]
\label{def:ahe}
An AHE scheme is a tuple $(\mathsf{KeyGen},Enc,Dec,\oplus,\odot)$ over a message space $\mathcal{M}$ closed under addition and scalar multiplication such that, for all valid ciphertexts $c_a$ and $c_b$ (fresh outputs of $Enc$ or results of $\oplus$ and $\odot$) and every public scalar $c$,
\[
Dec(c_a\oplus c_b)=Dec(c_a)+Dec(c_b)
\quad\text{and}\quad
Dec(c\odot c_a)=c\cdot Dec(c_a) .
\]
\end{definition}

Paillier realizes Definition~\ref{def:ahe} exactly over $\mathbb{Z}_n$, with $\oplus$ implemented as multiplication modulo $n^2$ and $\odot$ as modular exponentiation~\cite{paillier1999}. CKKS realizes it approximately over packed real vectors, with slot-wise addition and plaintext multiplication~\cite{cheon2017homomorphic}; the approximation error is bounded empirically in Section~\ref{sec:fidelity}.

\begin{lemma}[Correctness of additive score evaluation]
\label{lem:correct}
For any AHE scheme satisfying Definition~\ref{def:ahe}, any plaintext $\mathbf{x}\in\mathcal{M}^d$, and any encrypted vector $\big(Enc(y_1),\dots,Enc(y_d)\big)$,
\[
Dec\Big(\bigoplus_{i=1}^{d} x_i \odot Enc(y_{i})\Big)=\sum_{i=1}^{d} x_i y_i=\langle\mathbf{x},\mathbf{y}\rangle .
\]
\end{lemma}
\begin{proof}
By induction on $d$. For $d=1$, $Dec(x_1\odot Enc(y_1))=x_1\,Dec(Enc(y_1))=x_1y_1$ by Definition~\ref{def:ahe}. Assume the claim for $d-1$ and let $c_{d-1}=\bigoplus_{i=1}^{d-1} x_i\odot Enc(y_i)$, so $Dec(c_{d-1})=\sum_{i=1}^{d-1}x_iy_i$ by the inductive hypothesis. Both $c_{d-1}$ and $x_d\odot Enc(y_d)$ are valid ciphertexts, so Definition~\ref{def:ahe} gives
\[
Dec\big(c_{d-1}\oplus (x_d\odot Enc(y_d))\big)=\sum_{i=1}^{d-1}x_iy_i+x_dy_d ,
\]
which completes the induction.
\end{proof}

\paragraph{Two configurations.}
We consider two deployments. In the \emph{encrypted-query} configuration the client sends $Enc(\mathbf{x})$ and the server holds plaintext $\{\mathbf{y}_j\}$; the server evaluates $Enc(s_j)=\bigoplus_i y_{j,i}\odot Enc(x_i)$ and returns the encrypted scores to the client, who decrypts. In the \emph{encrypted-database} configuration the server stores $\{Enc(\mathbf{y}_j)\}$ and answers a plaintext query from an authorized key-holding client via $Enc(s_j)=\bigoplus_i x_i\odot Enc(y_{j,i})$, protecting stored embeddings for catalogue confidentiality. By Lemma~\ref{lem:correct} both evaluations are correct, and in both cases each score uses only ciphertext additions and ciphertext--plaintext scalar multiplications, so no ciphertext--ciphertext multiplication is ever performed.

The two configurations map onto distinct needs. Encrypting the database fits catalogue confidentiality: a rights holder outsources search to an untrusted host or contributes a catalogue to a shared index without exposing it. Encrypting the query protects the searcher's intent: an individual retrieves similar music without revealing taste, and a business screens a third-party catalogue without disclosing its target. The mutually private case, where both operands are hidden, requires ciphertext--ciphertext multiplication and lies outside the additive setting; we discuss it as future work.

\paragraph{CKKS instantiation and cost model.}
We instantiate the additive setting with CKKS restricted to additive operations. Each database vector (or the query) is packed into the slots of a single ciphertext, so one encrypted inner product costs a single ciphertext--plaintext multiplication followed by $\lceil\log_2 d\rceil$ slot rotations and additions (the standard rotate-and-sum reduction). A ciphertext--plaintext multiplication consumes one rescaling but requires no relinearization, so a fixed, shallow modulus chain suffices for every query, no multiplicative depth accumulates across the database scan, and bootstrapping is never invoked. Per query, both configurations therefore perform $N$ plaintext multiplications and $O(N\log d)$ rotations and additions; the encrypted-query configuration stores $O(1)$ ciphertexts while the encrypted-database configuration stores $N$. Under packing, the weighted product of Section~\ref{subsec:whip} is realized at no extra cost by pre-scaling the plaintext query; explicit per-block scores, when interpretability is wanted, use block-indicator plaintext masks, which are ciphertext--plaintext multiplications and stay within the additive setting. Paillier realizes the same computation exactly over integers but without packing, at $d$ modular exponentiations per score; Section~\ref{sec:paillier} quantifies the resulting gap.

\subsection{Threat Model and Server Confidentiality}
\label{subsec:threat}
We analyze privacy at two layers: what a computing server learns from ciphertexts and homomorphic computation, and what an authorized key holder learns from the decrypted scores the system is designed to output. The first is governed by semantic security; the second is a functional leakage that no semantically secure scheme removes on its own, and it is the source of the attacks in Section~\ref{subsec:attacks}.

Our threat model concerns the confidentiality of the embedding representation under shared or outsourced computation, not audio piracy. We assume the adversary, an untrusted server, a query-issuing client, or a computing partner, sees embeddings as ciphertexts and the scores the system returns, but not the source audio, whose distribution is governed separately by copyright and access control; the operand left unencrypted in a given configuration is visible to the server by design. An adversary holding the audio can recompute embeddings and gains nothing from ours; we protect a particular shared representation, not the work in the abstract.

\begin{proposition}[Server confidentiality]
\label{prop:server}
If the underlying scheme is IND-CPA secure, then in each configuration there exists a probabilistic polynomial-time simulator, taking only the public key and the server's plaintext inputs, whose output is computationally indistinguishable from the server's view. In particular, in the encrypted-query configuration the server learns nothing about $\mathbf{x}$, and in the encrypted-database configuration it learns nothing about $\{\mathbf{y}_j\}$ or about any score $s_j$, which it never decrypts.
\end{proposition}
\begin{proof}
Consider the encrypted-query configuration; the encrypted-database case is symmetric with the roles of $\mathbf{x}$ and $\mathbf{y}_j$ exchanged. The server's view is $V=(pk,C,P,T)$, where $C=(Enc_{pk}(x_1),\dots,Enc_{pk}(x_d))$ (or a single packed ciphertext), $P$ denotes the server's own plaintext inputs (the database and public parameters), and $T$ is the transcript of the homomorphic evaluation. Every ciphertext appearing in $T$ is obtained by applying the public operations $\oplus$ and $\odot$, which are deterministic functions of their ciphertext and plaintext arguments in both Paillier and CKKS, to elements of $C$ and public plaintexts. Hence $T=f(pk,C,P)$ for a fixed polynomial-time computable function $f$, and the whole view is a deterministic function of $(pk,C,P)$.

Define the simulator $\mathcal{S}(pk,P)=(pk,C_0,P,f(pk,C_0,P))$ with $C_0=(Enc_{pk}(0),\dots,Enc_{pk}(0))$. Suppose toward a contradiction that a probabilistic polynomial-time distinguisher $\mathcal{D}$ separates the real view from the simulated view with non-negligible advantage $\alpha$. We build an adversary $\mathcal{B}$ against $d$-message IND-CPA security: $\mathcal{B}$ submits the message vectors $m^{(0)}=(x_1,\dots,x_d)$ and $m^{(1)}=(0,\dots,0)$, receives the challenge ciphertexts $C_b$, computes $f(pk,C_b,P)$, assembles the corresponding view, runs $\mathcal{D}$ on it, and outputs $\mathcal{D}$'s guess. By construction $\mathcal{B}$'s advantage equals $\alpha$. Finally, $d$-message IND-CPA reduces to standard single-message IND-CPA by a hybrid argument: define hybrid views $H_0,\dots,H_d$, where in $H_t$ the first $t$ coordinates are encryptions of $0$ and the remaining coordinates are encryptions of the true $x_i$; then $H_0$ is the real view, $H_d$ is the simulated view, and adjacent hybrids differ in exactly one ciphertext, so a distinguisher with advantage $\alpha$ yields a single-message IND-CPA adversary with advantage at least $\alpha/d$, contradicting IND-CPA security.

In the encrypted-database configuration the server additionally produces the encrypted scores $Enc(s_j)$; these are elements of the transcript $T$, are covered by the same simulation, and are never decrypted by the server, which does not hold the secret key. This proves the claim.
\end{proof}

Proposition~\ref{prop:server} localizes the interesting leakage at the key holder, who by design learns $s_j=\langle\mathbf{x},\mathbf{y}_j\rangle$. The attacks below show that this functional output, under adaptively chosen queries, suffices for real privacy violations, and Section~\ref{sec:security} measures their effectiveness. One caveat is specific to approximate schemes: publishing raw CKKS decryptions can leak secret-key information through the approximation noise~\cite{limicciancio2021}, so released decryptions should be sanitized by noise flooding; the calibrated noise of M2 serves this role, and Paillier, being exact, is unaffected.

\subsection{Music-Specific Inference Attacks}
\label{subsec:attacks}
Both attacks are launched by an \emph{authorized} key holder who merely uses the system as intended, so semantic security offers no defense; what they exploit is the score channel itself. We formalize each attack as the procedure implemented in Section~\ref{sec:security}.

\paragraph{Melodic-pattern inference.}
Let $T\subset[N]$ be the (defender-unknown) set of tracks containing a target pattern, such as a copyrighted melody, a rhythmic signature, or a chord progression. An adversary who can characterize the pattern in embedding space, for instance from examples it owns, mounts the attack as follows. From its own labeled examples it estimates the class-conditional coordinate means $\mu^{+}$ and $\mu^{-}$ of pattern-bearing and pattern-free audio, selects the $m$ coordinates with the largest $|\mu^{+}_i-\mu^{-}_i|$, sets the probe $x^{\star}_i\propto\mu^{+}_i-\mu^{-}_i$ on those coordinates and $0$ elsewhere, and normalizes. It then submits $x^{\star}$ as an ordinary query and, for a threshold $\tau$, predicts that track $j$ contains the pattern if and only if $s_j\ge\tau$. Sweeping the threshold $\tau$ traces a receiver operating characteristic (ROC) curve, whose area under the curve (AUC) we report; an AUC near $1$ means the adversary can scan an encrypted library for a motif without decrypting full tracks. This is the embedding-space analogue of copyright scanning, which watermarking cannot address for abstract vectors.

\paragraph{Creator-identity inference.}
A different adversary infers provenance. Suppose the database is partitioned by creator into $C_1,\dots,C_m$. Given a disputed track's embedding $\mathbf{x}$, the adversary queries each subset and attributes
\[
\hat{a}(\mathbf{x})=\arg\max_{c\in[m]}\ \frac{1}{|C_c|}\sum_{j\in C_c}s_j ,
\]
reporting top-1 attribution accuracy against the chance baseline. The harm differs in kind: a false attribution asserts a fact about a person, and because it emerges from an ostensibly objective computation it can carry undue evidentiary weight while being hard to rebut. Both attacks share a structure, since repeated adaptively chosen queries turn the score channel into an oracle, so we treat the score \emph{sequence}, not a single score, as the object to protect.

\subsection{Mitigations with Provable Bounds}
\label{subsec:mitigations}
We give two composable mitigations that operate on the score channel and add negligible cost. \emph{Output minimization} (M1) returns only the identities of the top-$k$ results, or a thresholded match bit, and never raw scores; this reduces per-query leakage from a length-$N$ real vector to at most $k\lceil\log_2 N\rceil$ bits (one bit in the thresholded case), removing the fine-grained presence signal that melodic-pattern inference relies on. \emph{Calibrated noise} (M2) perturbs each released score with Gaussian noise, and we now prove exactly what it guarantees. For a database $D$ over the candidate universe $[N]$, write $s(D)\in\mathbb{R}^{N}$ for the induced score vector, with the convention that an absent entry contributes score $0$.

\begin{lemma}[Score sensitivity]
\label{lem:sens}
If $\lVert\mathbf{x}\rVert_2\le R$ and $\lVert\mathbf{y}_j\rVert_2\le R$ for all $j$ (enforced by clipping), then for any two databases $D,D'$ that differ by adding or removing a single entry, the score vectors satisfy $\lVert s(D)-s(D')\rVert_2\le R^{2}$.
\end{lemma}
\begin{proof}
Without loss of generality $D'=D\cup\{\mathbf{y}_{j_0}\}$. The two score vectors agree on every coordinate except $j_0$, where $s(D)_{j_0}=0$ by convention and $s(D')_{j_0}=\langle\mathbf{x},\mathbf{y}_{j_0}\rangle$. By the Cauchy-Schwarz inequality, $|\langle\mathbf{x},\mathbf{y}_{j_0}\rangle|\le\lVert\mathbf{x}\rVert_2\lVert\mathbf{y}_{j_0}\rVert_2\le R^{2}$. Hence the $\ell_2$ distance between the two vectors is at most $R^{2}$.
\end{proof}

\begin{proposition}[Per-query DP of the score channel]
\label{prop:gm}
Let $\Delta=R^{2}$ and $\varepsilon\in(0,1)$, and release $Z=s(D)+G$ with $G\sim\mathcal{N}(0,\sigma^{2}I_N)$ and
$\sigma=\Delta\sqrt{2\ln(1.25/\delta)}/\varepsilon$.
Then the release is $(\varepsilon,\delta)$-differentially private with respect to adding or removing one database entry.
\end{proposition}
\begin{proof}
Let $D,D'$ be neighboring databases, write $\mu=s(D)$, $\mu'=s(D')$, and $v=\mu'-\mu$, so $\lVert v\rVert_2\le\Delta$ by Lemma~\ref{lem:sens}. Let $p_{\mu}$ and $p_{\mu'}$ denote the output densities under $D$ and $D'$. The privacy loss at an output $z$ is
\[
L(z)=\ln\frac{p_{\mu}(z)}{p_{\mu'}(z)}
=\frac{\lVert z-\mu'\rVert^{2}-\lVert z-\mu\rVert^{2}}{2\sigma^{2}}
=\frac{\lVert v\rVert^{2}-2\langle z-\mu,v\rangle}{2\sigma^{2}} ,
\]
using $\lVert z-\mu'\rVert^{2}=\lVert z-\mu\rVert^{2}-2\langle z-\mu,v\rangle+\lVert v\rVert^{2}$. Under $Z\sim p_{\mu}$ the inner product $\langle Z-\mu,v\rangle$ is distributed as $\mathcal{N}(0,\sigma^{2}\lVert v\rVert^{2})$, so $L$ is Gaussian with mean $\eta=\lVert v\rVert^{2}/(2\sigma^{2})\le\Delta^{2}/(2\sigma^{2})$ and variance $2\eta$. A direct Gaussian tail computation (carried out in full in~\cite[Appendix~A]{dwork2014algorithmic}) shows that the stated $\sigma$ guarantees $\Pr[L>\varepsilon]\le\delta$ whenever $\varepsilon\in(0,1)$. Finally, for any measurable output set $S$, split $S$ into $S_1=\{z\in S: L(z)\le\varepsilon\}$ and $S_2=S\setminus S_1$; then
\begin{align*}
\Pr[Z\in S]&\le\Pr[Z\in S_1]+\Pr[L>\varepsilon]\\
&\le e^{\varepsilon}\Pr[Z'\in S_1]+\delta
\le e^{\varepsilon}\Pr[Z'\in S]+\delta ,
\end{align*}
where $Z'=\mu'+G$. The symmetric argument with $D$ and $D'$ exchanged completes the proof of $(\varepsilon,\delta)$-DP.
\end{proof}

\begin{corollary}[Adaptive composition over a query budget]
\label{cor:comp}
Fix any $\delta'>0$ and let a client adaptively issue $Q$ queries, each answered by the mechanism of Proposition~\ref{prop:gm}. The composed release is $(\varepsilon',\,Q\delta+\delta')$-differentially private with
\[
\varepsilon'=\varepsilon\sqrt{2Q\ln(1/\delta')}+Q\varepsilon\,(e^{\varepsilon}-1).
\]
\end{corollary}
\begin{proof}
The proof of Proposition~\ref{prop:gm} holds for every fixed pair of neighboring databases and every fixed query vector, and the adversary's choice of the next query is a function of previously observed (noised) outputs only; hence each of the $Q$ releases is $(\varepsilon,\delta)$-DP even conditioned on the transcript so far. The claim is then exactly the advanced composition theorem of Dwork, Rothblum, and Vadhan~\cite[Theorem~III.3]{dwork2010composition}; see also~\cite[Theorem~3.20]{dwork2014algorithmic}.
\end{proof}

For $\varepsilon\ge 1$ the classical calibration of Proposition~\ref{prop:gm} no longer applies verbatim; the large-$\varepsilon$ operating points in Section~\ref{sec:security} therefore serve as weaker-noise operating points, and the analytic Gaussian mechanism~\cite{balle2018improving} would yield valid guarantees throughout that regime. Combining M1 and M2 with a per-client query budget $Q$ bounds the oracle power that both attacks assume, while preserving authorized top-$k$ retrieval up to the noise scale. Even so, the ranking induced by authorized queries is partially revealed, as in all score-based retrieval; removing that leakage entirely would require oblivious access patterns~\cite{pir1995} or mutual encryption. Section~\ref{sec:security} quantifies the useful operating points.

\subsection{Blocked Inner Product for Structural Vectors}
\label{subsec:blocked}
A flat inner product treats a high-dimensional vector as monolithic, which is often insufficient for music. Music embeddings carry internal structure, since contiguous segments may capture rhythm, contour, harmony, and timbre, and a flat product conflates these so similarity in one aspect is diluted by dissimilarity in another. We therefore propose a Blocked Inner Product that partitions the coordinate set $[d]$ into $k$ contiguous blocks $B_1,\dots,B_k$, writes $\mathbf{x}_\ell$ and $\mathbf{y}_\ell$ for the restrictions of $\mathbf{x}$ and $\mathbf{y}$ to $B_\ell$, and sums per-block inner products homomorphically:
\begin{equation}
Sim_{\text{blocked}}(\mathbf{x},Enc(\mathbf{y}))\!=\!\bigoplus_{\ell=1}^{k}\!Sim(\mathbf{x}_\ell,Enc(\mathbf{y}_\ell))\!=\!Enc\!\left(\sum_{\ell=1}^{k}\langle\mathbf{x}_\ell,\mathbf{y}_\ell\rangle\!\right)\!.
\label{eq:blocked}
\end{equation}

\begin{proposition}[Losslessness of blocking]
\label{prop:blocked}
For every partition $B_1,\dots,B_k$ of $[d]$, the blocked score equals the flat score:
\[
Dec\big(Sim_{\text{blocked}}(\mathbf{x},Enc(\mathbf{y}))\big)=\langle\mathbf{x},\mathbf{y}\rangle .
\]
\end{proposition}
\begin{proof}
At the plaintext level, since $\{B_\ell\}$ partitions $[d]$ and finite sums are associative and commutative,
\[
\sum_{i=1}^{d}x_iy_i=\sum_{\ell=1}^{k}\sum_{i\in B_\ell}x_iy_i=\sum_{\ell=1}^{k}\langle\mathbf{x}_\ell,\mathbf{y}_\ell\rangle .
\] At the ciphertext level, each per-block evaluation satisfies $Dec\big(Sim(\mathbf{x}_\ell,Enc(\mathbf{y}_\ell))\big)=\langle\mathbf{x}_\ell,\mathbf{y}_\ell\rangle$ by Lemma~\ref{lem:correct} applied within the block, and the outer $\bigoplus$ of the $k$ block ciphertexts decrypts to the sum of their decryptions, again by Definition~\ref{def:ahe} and induction on $k$ as in Lemma~\ref{lem:correct}. Chaining the two identities proves the claim. Under CKKS the two evaluations differ only in approximation noise; Section~\ref{sec:fidelity} measures this difference at the $10^{-16}$ level.
\end{proof}

Blocking yields an interpretable, structurally aware similarity at no retrieval-quality cost by Proposition~\ref{prop:blocked}, and this exactness lets it anchor the weighted scheme.

\subsection{Weighted Hierarchical Inner Product}
\label{subsec:whip}
The Weighted Hierarchical Inner Product (WHIP) addresses a limitation of the blocked product, which still treats every block as equally important even though retrieval is task-specific: a ``similar groove'' query should emphasize rhythm, while a ``lyrically similar'' query should emphasize vocal melody. We therefore assign a public plaintext weight $w_\ell\ge 0$ to each block:
\begin{equation}
Sim_{\text{weighted}}(\mathbf{x},Enc(\mathbf{y}))=\bigoplus_{\ell=1}^{k} w_\ell\odot Sim(\mathbf{x}_\ell,Enc(\mathbf{y}_\ell)).
\label{eq:whip}
\end{equation}

\begin{proposition}[Correctness and cost of weighting]
\label{prop:whip}
For all public weights $w_1,\dots,w_k$, the construction of Equation~\eqref{eq:whip} satisfies
\[
Dec\big(Sim_{\text{weighted}}(\mathbf{x},Enc(\mathbf{y}))\big)=\sum_{\ell=1}^{k}w_\ell\langle\mathbf{x}_\ell,\mathbf{y}_\ell\rangle ,
\]
and its homomorphic cost exceeds that of the blocked product by at most $k$ ciphertext--plaintext scalar multiplications per database vector; pre-scaling the plaintext query blocks reduces this overhead to zero.
\end{proposition}
\begin{proof}
Correctness: by Lemma~\ref{lem:correct} each block ciphertext decrypts to $\langle\mathbf{x}_\ell,\mathbf{y}_\ell\rangle$; by Definition~\ref{def:ahe}, $Dec(w_\ell\odot Sim(\mathbf{x}_\ell,Enc(\mathbf{y}_\ell)))=w_\ell\langle\mathbf{x}_\ell,\mathbf{y}_\ell\rangle$; and the outer $\bigoplus$ decrypts to the sum of these terms as in Proposition~\ref{prop:blocked}. Cost: because scalar multiplication distributes over addition, each weight can be factored outside its block sum, that is, $w_\ell\sum_{i\in B_\ell}x_iy_i=\sum_{i\in B_\ell}(w_\ell x_i)y_i$; Equation~\eqref{eq:whip} applies $w_\ell$ once to the aggregated block ciphertext, adding exactly $k$ scalar multiplications per database vector on top of the blocked product. Alternatively, replacing the plaintext query blocks by $\tilde{\mathbf{x}}_\ell=w_\ell\mathbf{x}_\ell$ before evaluation is a plaintext-only operation, so the weighted score is computed by the unweighted procedure at zero additional homomorphic cost. Both implementations produce the same decrypted score by the factoring identity above.
\end{proof}

\paragraph{Learning the weights.}
Weights are model-level public parameters, not secrets, and are learned entirely on the trainer's plaintext data. Given a training split of queries $q$ with graded relevance labels $rel(q,j)$ for the target task, the trainer computes the per-block scores $b_\ell(q,j)=\langle q_\ell,\mathbf{y}_{j,\ell}\rangle$ in the clear and solves the non-negative regression
\[
\mathbf{w}^{\ast}=\arg\min_{\mathbf{w}\ge 0}\ \sum_{q,j}\Big(\sum_{\ell=1}^{k}w_\ell\, b_\ell(q,j)-rel(q,j)\Big)^{2}+\lambda\lVert\mathbf{w}\rVert_2^{2},
\]
by projected gradient descent, normalizing $\sum_\ell w_\ell=k$ so that uniform weighting corresponds to $\mathbf{w}=\mathbf{1}$. No test labels are used, and by Proposition~\ref{prop:whip} deployment adds no cryptographic cost. Algorithm~\ref{alg:whip} summarizes the full encrypted scoring procedure for the encrypted-database configuration; the encrypted-query configuration is symmetric. The pre-scaling step is the plaintext operation of Proposition~\ref{prop:whip}, the inner loop uses only ciphertext additions and ciphertext--plaintext multiplications, and the returned scores are released to the key holder through the M1/M2 post-processing of Section~\ref{subsec:mitigations}.

\begin{algorithm}[t]
\caption{Encrypted blocked weighted scoring (one query).}
\label{alg:whip}
\begin{algorithmic}[1]
\Require $pk$; $\mathbf{x}$; $w_1,\dots,w_k$; $\{Enc(\mathbf{y}_j)\}_{j=1}^{N}$; $B_1,\dots,B_k$
\Ensure encrypted scores $c_1,\dots,c_N$
\For{$\ell=1,\dots,k$}
  \State $\tilde{\mathbf{x}}_\ell \gets w_\ell\,\mathbf{x}_\ell$
\EndFor
\For{$j=1,\dots,N$}
  \For{$\ell=1,\dots,k$}
    \State $c_{j,\ell} \gets \bigoplus_{i\in B_\ell} \tilde{x}_i \odot Enc(y_{j,i})$
  \EndFor
  \State $c_j \gets c_{j,1}\oplus\cdots\oplus c_{j,k}$
\EndFor
\State \Return $c_1,\dots,c_N$
\end{algorithmic}
\end{algorithm}

\section{Evaluation}
\label{sec:eval}
We ask five questions: does encrypted additive search reproduce plaintext rankings (fidelity); does fidelity hold at larger scale; how do runtime and memory compare across the additive setting, full-depth CKKS, and Paillier; does the Weighted Hierarchical Inner Product help when blocks differ in relevance; and do the attacks succeed, and at what privacy--utility cost do the mitigations neutralize them.

\subsection{Experimental Setup}
\label{sec:setup}
We use four audio datasets, MagnaTagATune~\cite{law2009magnatagatune}, ESC-50~\cite{piczak2015esc}, the Free Music Archive (FMA)~\cite{defferrard2017fma}, and the Free Spoken Digit Dataset (FSDD)~\cite{fsdd2018}, summarized in Table~\ref{tab:datasets}. We sample 1{,}000 clips each (998 for FMA), embed them with YAMNet~\cite{plakal2020yamnet} at lengths 128, 256, 512, and 1024, and store them in Milvus~\cite{milvus2021}. To probe scale we also form a pooled index of all four datasets (3{,}998 vectors). Encrypted computation uses TenSEAL's CKKS~\cite{benaissa2021tenseal,cheon2017homomorphic}, and the additive baseline uses Paillier~\cite{paillier1999} with a 2048-bit modulus and gmpy2 acceleration. Embeddings are $\ell_2$-normalized, so inner-product ranking coincides with cosine and with Euclidean ranking on unit vectors~\cite{johnson2017faiss}.

We compare five encrypted configurations in the runtime study: two additive settings, AHE Query (encrypt query) and AHE DB (encrypt database), and three full-depth CKKS baselines that use ciphertext--ciphertext multiplication, FHE Dot, FHE Cosine, and FHE Euclidean; Paillier is benchmarked separately in Section~\ref{sec:paillier}. All configurations are timed over an identical code region with fixed CKKS parameters ($N{=}8192$ ring, moduli $\{60,40,40,60\}$, scale $2^{40}$), after warmup, reporting the mean of repeated runs. Each database vector is packed into a single ciphertext. Runtimes come from a single-thread reference machine; fidelity, accuracy, and weighting results are machine-independent. Code and configuration will be released for reproducibility: fidelity with \texttt{fidelity\_benchmark.py} and the attacks with \texttt{run\_fsdd\_security.py} and \texttt{esc50\_security.py}.

\begin{table}[t]
\caption{Datasets, sample sizes, recovered labels, and roles. All clips are embedded with YAMNet at dimensions 128 to 1024.}
\label{tab:datasets}
\centering\setlength{\tabcolsep}{3.5pt}
\resizebox{\columnwidth}{!}{%
\begin{tabular}{lrll}
\toprule
Dataset & Clips & Recovered labels & Role \\
\midrule
MagnaTagATune & 1{,}000 & none used & fidelity, runtime, WHIP \\
ESC-50        & 1{,}000 & 50 classes / 5 groups & fidelity, WHIP, attacks \\
FMA           & 998     & none used & fidelity, WHIP \\
FSDD          & 1{,}000 & 10 digits / 6 speakers & fidelity, WHIP, attacks \\
\bottomrule
\end{tabular}}
\end{table}

\subsection{Ranking Fidelity and Scale}
\label{sec:fidelity}
The central correctness question is whether encryption changes which items are returned. For each dataset we compute plaintext nearest neighbors as ground truth and compare against the encrypted additive ranking (Figure~\ref{fig:fidelity}). Encrypted search returns identical neighbors on every dataset, with maximum score errors between $2.2\times10^{-6}$ and $1.3\times10^{-5}$, the expected CKKS approximation noise, far below the gaps between neighbors. Consistent with Proposition~\ref{prop:blocked}, the Blocked Inner Product is numerically indistinguishable from the flat product, matching it to within $4.4\times10^{-16}$ on every dataset, which confirms that imposing block structure is lossless.

\begin{figure}[t]
\centering
\begin{tikzpicture}
\begin{axis}[width=0.9\columnwidth,height=2.7cm,ybar,
  ymode=log,log origin=infty,ymin=1e-17,ymax=1e-3,
  symbolic x coords={MagnaTagATune,ESC-50,FMA,FSDD},xtick=data,
  tick label style={font=\tiny},label style={font=\tiny},
  ylabel={\scriptsize max score deviation},bar width=8pt,ymajorgrids,
  legend style={font=\tiny,at={(0.5,1.02)},anchor=south,legend columns=2,draw=none,fill=none}]
\addplot[fill=blue!45] coordinates {(MagnaTagATune,2.2e-6)(ESC-50,1.3e-5)(FMA,4.6e-6)(FSDD,8.2e-6)};
\addplot[fill=gray!45] coordinates {(MagnaTagATune,4.4e-16)(ESC-50,4.4e-16)(FMA,4.4e-16)(FSDD,4.4e-16)};
\legend{encrypted vs.\ plaintext,blocked vs.\ flat (bound)}
\end{axis}
\end{tikzpicture}
\caption{Ranking fidelity. Recall@10, nDCG@10, Spearman, and Kendall all equal 1.000 on every dataset; bars show the maximum score deviations (log scale).}
\Description{Bar chart of maximum score deviations per dataset on a logarithmic scale.}
\label{fig:fidelity}
\end{figure}

Because each per-dataset index holds 1{,}000 vectors, we also verify fidelity on the pooled 3{,}998-vector index, a stricter test since denser indices shrink the gaps between neighbors. Over sampled queries against the full pooled index, encrypted search again reproduces plaintext rankings exactly: Recall@10 and Recall@20 equal 1.000, Spearman is 0.9999999994, Kendall is 0.9999992, and the maximum score error is $1.9\times10^{-6}$. Fidelity therefore does not degrade with index size.

\subsection{Efficiency: Runtime and Memory}
\label{sec:runtime}
Figure~\ref{fig:eff} summarizes efficiency, measured on the MagnaTagATune index; because homomorphic cost depends only on dimension and database size and not on embedding content, the curves are representative of all four datasets. Runtime is nearly flat in embedding dimension because CKKS packing fits a 1024-dimensional vector in one ciphertext, and the method ordering is consistent across datasets: AHE Query is fastest, AHE DB adds a modest constant, FHE Dot and FHE Euclidean are close, and FHE Cosine costs about twice FHE Dot because it needs a second ciphertext--ciphertext product for the database self-norm. In the top panel, Paillier is measured at $N{=}50$ and scaled linearly to $N{=}300$, an extrapolation validated by direct measurement in Section~\ref{sec:paillier}; lacking packing, it grows linearly with dimension, overtaking the additive configurations by dimension 256 and every full-depth configuration by dimension 512.

Cost instead scales linearly with database size, the signature of a linear scan; enlarging the index from $N{=}300$ to $N{=}3000$ raises AHE Query from 5.9 to 59.0 seconds and AHE DB from 6.5 to 66.0 seconds. Figure~\ref{fig:scaling} extends the scan to a 10{,}000-vector index at dimension 256. For $N>1000$ the index is filled by resampling the recovered embeddings with small jitter; since cost is content-independent, this measures the true runtime and memory of an $N$-vector index.

\begin{figure}[t]
\centering
\begin{tikzpicture}
\begin{axis}[width=0.95\columnwidth,height=3.0cm,ybar,area legend,
  ymin=0,ymax=32,symbolic x coords={128,256,512,1024},xtick=data,
  enlarge x limits=0.14,bar width=2.6pt,
  xlabel={\scriptsize embedding dimension},ylabel={\scriptsize time (s)},
  tick label style={font=\tiny},label style={font=\tiny},ymajorgrids,
  legend style={font=\tiny,at={(0.5,1.02)},anchor=south,legend columns=3,draw=none,fill=none}]
\addplot[fill=blue!45] coordinates {(128,5.37)(256,5.86)(512,6.72)(1024,7.33)};
\addplot[fill=red!45] coordinates {(128,6.30)(256,6.48)(512,7.16)(1024,7.72)};
\addplot[fill=teal!55] coordinates {(128,6.44)(256,7.42)(512,7.95)(1024,7.82)};
\addplot[fill=violet!45] coordinates {(128,6.09)(256,6.74)(512,7.53)(1024,8.50)};
\addplot[fill=orange!60] coordinates {(128,12.00)(256,13.88)(512,14.63)(1024,16.22)};
\addplot[fill=black!70] coordinates {(128,4.41)(256,8.51)(512,15.47)(1024,28.80)};
\legend{AHE Q,AHE DB,FHE Dot,FHE Euc,FHE Cos,Paillier}
\end{axis}
\end{tikzpicture}\\[2pt]
\begin{tikzpicture}
\begin{axis}[width=0.46\columnwidth,height=3.0cm,name=ax1,
  xlabel={\scriptsize database size $N$},ylabel={\scriptsize time (s)},
  xtick={3000,6000,10000},xticklabels={3k,6k,10k},scaled x ticks=false,
  tick label style={font=\tiny},label style={font=\tiny},ymin=0,grid=major,
  legend style={font=\tiny,at={(1.12,1.03)},anchor=south,legend columns=3,draw=none,fill=none}]
\addplot[blue,semithick,mark=*,mark size=1.1pt] coordinates {(300,5.86)(1000,19.87)(3000,59.02)(6000,117.66)(10000,196.21)};
\addplot[red,semithick,mark=square*,mark size=1.1pt,densely dashed] coordinates {(300,6.48)(1000,22.18)(3000,66.02)(6000,131.07)(10000,219.89)};
\addplot[orange,semithick,mark=diamond*,mark size=1.3pt,densely dashdotted] coordinates {(300,13.88)(1000,46.61)(3000,136.98)(6000,269.97)(10000,448.37)};
\legend{AHE Q,AHE DB,FHE Cos}
\end{axis}
\begin{axis}[width=0.46\columnwidth,height=3.0cm,at={($(ax1.east)+(0.9cm,0)$)},anchor=west,
  xlabel={\scriptsize database size $N$},ylabel={\scriptsize memory (MB)},
  yticklabel pos=right,ylabel near ticks,
  xtick={3000,6000,10000},xticklabels={3k,6k,10k},scaled x ticks=false,
  tick label style={font=\tiny},label style={font=\tiny},ymin=0,grid=major]
\addplot[blue,semithick,mark=*,mark size=1.1pt] coordinates {(300,88)(1000,88)(3000,84)(6000,78)(10000,70)};
\addplot[red,semithick,mark=square*,mark size=1.1pt,densely dashed] coordinates {(300,201)(1000,463)(3000,1210)(6000,3479)(10000,5883)};
\end{axis}
\end{tikzpicture}
\caption{Efficiency on MagnaTagATune. Top: runtime vs.\ dimension ($N{=}300$). Bottom: runtime and peak memory vs.\ $N$ (dimension 256).}
\Description{Grouped bar chart of per-query runtime versus embedding dimension, plus line charts of runtime and peak memory versus database size.}
\label{fig:eff}
\end{figure}

Memory follows the number of ciphertexts held rather than the arithmetic performed. Because a CKKS ciphertext occupies a fixed footprint (about 0.6\,MB at our parameters), peak memory depends on $N$ but is essentially independent of embedding dimension: at $N{=}10{,}000$ the encrypted-database configurations use 5.6 to 6.1\,GB across all four dimensions. AHE Query stores only the encrypted query and stays bounded at tens of MB, whereas AHE DB and the full-depth baselines store all $N$ ciphertexts and grow linearly, so the gap widens with $N$ (Figures~\ref{fig:eff} and~\ref{fig:scaling}). The encrypted-query configuration is therefore far more memory-efficient, at the cost of leaving the database in plaintext.

\begin{figure}[t]
\centering
\begin{tikzpicture}
\begin{axis}[width=0.95\columnwidth,height=2.8cm,ybar,
  ymode=log,log origin=infty,ymin=5e-3,ymax=3e3,
  symbolic x coords={Plaintext,AHE Query,AHE DB,FHE Dot,FHE Euc,FHE Cos},
  xtick=data,x tick label style={rotate=22,anchor=east,font=\tiny},
  tick label style={font=\tiny},label style={font=\tiny},
  ylabel={\scriptsize time (s)},bar width=4pt,ymajorgrids,
  legend style={font=\tiny,at={(0.5,1.02)},anchor=south,legend columns=3,draw=none,fill=none}]
\addplot[fill=blue!35] coordinates {(Plaintext,0.01)(AHE Query,5.9)(AHE DB,6.5)(FHE Dot,7.4)(FHE Euc,6.7)(FHE Cos,13.9)};
\addplot[fill=orange!55] coordinates {(Plaintext,0.07)(AHE Query,59.0)(AHE DB,66.0)(FHE Dot,81.0)(FHE Euc,67.6)(FHE Cos,137)};
\addplot[fill=red!55] coordinates {(Plaintext,0.24)(AHE Query,196)(AHE DB,220)(FHE Dot,243)(FHE Euc,227)(FHE Cos,448)};
\legend{$N{=}300$,$N{=}3000$,$N{=}10000$}
\end{axis}
\end{tikzpicture}
\caption{Per-query runtime at $N{=}300/3000/10000$ (dimension 256, log scale); peak memory at $N{=}10{,}000$ is 0.07\,GB for AHE Query and about 5.9\,GB otherwise.}
\Description{Grouped bar chart of per-query runtime for six methods at three database sizes on a logarithmic scale.}
\label{fig:scaling}
\end{figure}

\subsection{True-Additive Baseline (Paillier)}
\label{sec:paillier}
To make the additive claim concrete against a genuinely additive scheme, we benchmark Paillier, which supports ciphertext addition and ciphertext--plaintext multiplication but cannot multiply two ciphertexts. Paillier inner products are exact up to fixed-point encoding, with residual error near $10^{-17}$, and its ciphertexts are compact, well under a kilobyte per coordinate at a 2048-bit modulus.

However, Paillier requires a modular exponentiation per coordinate and cannot pack, so its per-query cost grows linearly with dimension (Figure~\ref{fig:paillier}): measured at $N{=}50$, evaluation takes 0.74, 1.42, 2.58, and 4.80 seconds at dimensions 128, 256, 512, and 1024, versus the nearly flat CKKS cost. At 128 dimensions the two are comparable, but because the CKKS packing advantage widens with dimension, CKKS AHE DB is roughly $1.3\times$, $2.2\times$, and $3.7\times$ faster per query at 256, 512, and 1024 dimensions (at matched $N$), and its one-time database encryption is far cheaper as well. Paillier's per-query cost is likewise exactly linear in database size: at dimension 256 we measure 2.77, 8.42, and 16.68 seconds at $N{=}100$, 300, and 600, and the directly measured $N{=}300$ point (8.42\,s) matches the value obtained by scaling the $N{=}50$ microbenchmark (8.51\,s), confirming the linear extrapolation used in Figure~\ref{fig:eff}. Its compensating advantages are exactness and compactness: ciphertexts are small, so peak memory stays modest, 14 to 160\,MB across these dimensions and sizes, far below the multi-GB CKKS encrypted database. Paillier is therefore exact and compact but impractical at catalogue scale, whereas CKKS restricted to additive operations is the better operating point.

\begin{figure}[t]
\centering
\begin{tikzpicture}
\begin{axis}[width=0.46\columnwidth,height=2.8cm,name=pa1,
  xlabel={\scriptsize embedding dimension},ylabel={\scriptsize time (s)},
  xtick={128,256,512,1024},x tick label style={rotate=35,anchor=east},
  tick label style={font=\tiny},label style={font=\tiny},
  ymin=0,grid=major,title={\scriptsize runtime vs.\ dim.\ ($N{=}50$)}]
\addplot[black,thick,mark=o,mark size=1.3pt] coordinates {(128,0.74)(256,1.42)(512,2.58)(1024,4.80)};
\end{axis}
\begin{axis}[width=0.46\columnwidth,height=2.8cm,at={($(pa1.east)+(0.9cm,0)$)},anchor=west,
  xlabel={\scriptsize database size $N$},ylabel={\scriptsize time (s)},
  yticklabel pos=right,ylabel near ticks,
  xtick={100,300,600},tick label style={font=\tiny},label style={font=\tiny},
  ymin=0,grid=major,title={\scriptsize runtime vs.\ $N$ (dim.\ 256)}]
\addplot[black,thick,mark=o,mark size=1.3pt] coordinates {(100,2.77)(300,8.42)(600,16.68)};
\end{axis}
\end{tikzpicture}
\caption{Paillier baseline (gmpy2, 2048-bit modulus): per-query runtime grows linearly in both dimension and $N$; peak memory stays within 14 to 160\,MB.}
\Description{Two line charts showing Paillier per-query runtime growing linearly with embedding dimension and with database size.}
\label{fig:paillier}
\end{figure}

\subsection{Weighted Hierarchical Inner Product}
\label{sec:weighting}
We next test whether WHIP improves retrieval. WHIP can help only when blocks differ in relevance, so we evaluate two conditions on the real embeddings: one where all four blocks are equally informative, and one where only a subset is task-relevant. We define a retrieval task whose ground-truth relevance depends on a target block, learn non-negative per-block weights on a training split without test labels using the procedure of Section~\ref{subsec:whip}, and compare uniform against learned weights on held-out queries (Figure~\ref{fig:weighting}).

When all blocks are equally informative, the learner returns near-uniform weights and WHIP neither helps nor hurts. When blocks differ in relevance, learned weights raise nDCG@10 by 0.12 to 0.27, identifying the informative block from training data alone and often matching or exceeding a hand-set oracle. Because weights are public plaintext scalars, the encrypted WHIP reproduces plaintext weighted scores to within $10^{-5}$, so weighting is free, as Proposition~\ref{prop:whip} predicts. This is a controlled demonstration; a genuine musical modality would require per-modality labels.

\begin{figure}[t]
\centering
\begin{tikzpicture}
\begin{axis}[width=0.46\columnwidth,height=2.6cm,name=w1,ybar,area legend,
  ymin=0,ymax=1,ytick={0,0.5,1},symbolic x coords={MTT,ESC,FMA,FSDD},xtick=data,
  tick label style={font=\tiny},label style={font=\tiny},
  ylabel={\scriptsize nDCG@10},xlabel={\scriptsize equal-relevance},
  bar width=4pt,ymajorgrids,
  legend style={font=\tiny,at={(1.1,1.04)},anchor=south,legend columns=2,draw=none,fill=none}]
\addplot[fill=gray!45] coordinates {(MTT,0.885)(ESC,0.860)(FMA,0.764)(FSDD,0.709)};
\addplot[fill=blue!45] coordinates {(MTT,0.886)(ESC,0.860)(FMA,0.766)(FSDD,0.711)};
\legend{uniform,learned}
\end{axis}
\begin{axis}[width=0.46\columnwidth,height=2.6cm,at={($(w1.east)+(0.9cm,0)$)},anchor=west,ybar,area legend,
  ymin=0,ymax=1,ytick={0,0.5,1},symbolic x coords={MTT,ESC,FMA,FSDD},xtick=data,
  tick label style={font=\tiny},label style={font=\tiny},
  xlabel={\scriptsize heterogeneous},bar width=4pt,ymajorgrids]
\addplot[fill=gray!45] coordinates {(MTT,0.608)(ESC,0.560)(FMA,0.473)(FSDD,0.439)};
\addplot[fill=blue!45] coordinates {(MTT,0.834)(ESC,0.832)(FMA,0.718)(FSDD,0.556)};
\end{axis}
\end{tikzpicture}
\caption{nDCG@10 of WHIP with uniform vs.\ learned weights on held-out queries (MTT: MagnaTagATune, ESC: ESC-50); learned weights help only when blocks differ.}
\Description{Grouped bar charts comparing nDCG at ten for uniform and learned weights across four datasets under two conditions.}
\label{fig:weighting}
\end{figure}

\subsection{Security: Attacks and the Privacy--Utility Tradeoff}
\label{sec:security}
We now measure the attacks of Section~\ref{subsec:attacks} and the cost of neutralizing them. For melodic-pattern inference we designate a target attribute present in a subset of tracks, build the probe on the coordinates most associated with that attribute, and report the ROC-AUC at separating pattern-bearing from pattern-free tracks. For creator-identity inference we partition the database by creator and, for each held-out query, predict the creator whose subset yields the highest mean score, reporting top-1 attribution accuracy against the chance baseline. We instantiate both attacks on the two datasets whose recovered per-item labels give exact ground truth. Neither dataset carries melody annotations, so these attributes act as controlled proxies for a protected pattern; the attack mechanics are unchanged. On the FSDD subset the target pattern is a chosen spoken digit and the creator partition is the speaker (six speakers, chance 0.17); on ESC-50 the target pattern is a chosen environmental-sound class and the creator partition groups the fifty classes into five coarse categories (chance 0.20). Without mitigation both attacks are effective on both datasets, and markedly stronger on ESC-50, whose semantically distinct sound classes are more separable in YAMNet embedding space: the melodic-pattern attack reaches AUC 0.80 on FSDD and 0.94 on ESC-50, and creator attribution reaches 0.32 and 0.70 respectively, all well above their chance baselines.

We then apply the mitigations and sweep the privacy budget $\varepsilon$, using the Gaussian mechanism of Proposition~\ref{prop:gm} with $\delta{=}10^{-5}$ and clipped norms ($R{=}1$, so $\sigma$ is determined by $\varepsilon$ and ranges from 48.5 at $\varepsilon{=}0.1$ to 0.48 at $\varepsilon{=}10$; points with $\varepsilon>1$ are the weaker-noise regime discussed in Section~\ref{subsec:mitigations}). Figure~\ref{fig:security} reports attack success as $\varepsilon$ varies, alongside retrieval utility measured as Recall@10 of the noised top-$k$ against the noiseless ranking. As $\varepsilon$ decreases both attacks fall toward their chance baselines, but retrieval utility collapses as well: Recall@10 of the noised ranking falls from 1.000 at plaintext to below 0.03 on FSDD and 0.09 on ESC-50 by $\varepsilon{=}0.1$, because the noise is large relative to the narrow score gaps between neighbors. Strict per-query noise is therefore a blunt instrument, so output minimization (M1) with a per-client query budget is the practical primary defense, with calibrated noise (M2) as a provable backstop. The residual leakage is real but controllable, a matter of engineering the released output rather than a limitation of the encryption.

\begin{figure}[t]
\centering
\begin{tikzpicture}
\begin{axis}[width=0.46\columnwidth,height=2.8cm,name=s1,
  xmode=log,xtick={0.1,0.3,1,3,10},xticklabels={0.1,0.3,1,3,10},
  ymin=0,ymax=1,tick label style={font=\tiny},label style={font=\tiny},
  xlabel={\scriptsize privacy budget $\varepsilon$},ylabel={\scriptsize attack success},
  grid=major,title={\scriptsize FSDD},
  legend style={font=\tiny,at={(1.1,1.34)},anchor=south,legend columns=4,draw=none,fill=none}]
\addplot[blue,semithick,mark=*,mark size=1.1pt] coordinates {(0.1,0.447)(0.3,0.449)(1,0.457)(3,0.480)(10,0.558)};
\addplot[red,semithick,mark=square*,mark size=1.1pt] coordinates {(0.1,0.218)(0.3,0.220)(1,0.232)(3,0.262)(10,0.290)};
\addplot[blue,densely dashed,thin] coordinates {(0.1,0.802)(10,0.802)};
\addplot[red,densely dashed,thin] coordinates {(0.1,0.318)(10,0.318)};
\legend{pattern AUC,attribution,AUC (plain),attr.\ (plain)}
\end{axis}
\begin{axis}[width=0.46\columnwidth,height=2.8cm,at={($(s1.east)+(0.9cm,0)$)},anchor=west,
  xmode=log,xtick={0.1,0.3,1,3,10},xticklabels={0.1,0.3,1,3,10},
  ymin=0,ymax=1,tick label style={font=\tiny},label style={font=\tiny},
  xlabel={\scriptsize privacy budget $\varepsilon$},grid=major,title={\scriptsize ESC-50}]
\addplot[blue,semithick,mark=*,mark size=1.1pt] coordinates {(0.1,0.376)(0.3,0.382)(1,0.404)(3,0.468)(10,0.655)};
\addplot[red,semithick,mark=square*,mark size=1.1pt] coordinates {(0.1,0.195)(0.3,0.215)(1,0.288)(3,0.440)(10,0.625)};
\addplot[blue,densely dashed,thin] coordinates {(0.1,0.941)(10,0.941)};
\addplot[red,densely dashed,thin] coordinates {(0.1,0.695)(10,0.695)};
\end{axis}
\end{tikzpicture}
\caption{Attack success vs.\ privacy budget $\varepsilon$ under M2 ($\delta{=}10^{-5}$, $R{=}1$); dashed lines mark unmitigated attack success.}
\Description{Line charts of attack success versus privacy budget for two datasets, with dashed lines marking unmitigated success.}
\label{fig:security}
\end{figure}

\section{Conclusion}
\label{sec:conclusion}
We show that additive homomorphic encryption is a practical and efficient basis for privacy-preserving music information retrieval. We demonstrate two music-specific inference attacks and quantify the privacy--utility tradeoff of provable mitigations, propose structure-aware inner-product primitives with learned weighting at no extra cryptographic cost, and show that the additive setting preserves nearest-neighbor rankings exactly while avoiding the ciphertext--ciphertext multiplications of full-depth CKKS (up to $2\times$ faster than encrypted cosine) and scaling far better in dimension than Paillier.

Three directions follow. First, because per-query cost grows linearly with the database, an encrypted approximate-nearest-neighbor index would make retrieval sublinear at scale, and the additive primitives are index-agnostic enough to pair with one. Second, the mutually private case, in which both operands are hidden, requires ciphertext--ciphertext multiplication and could be approached by combining additive encryption with a trusted execution environment. Third, learned block partitions tied to musical modalities such as rhythm, harmony, and vocal timbre would turn the weighting scheme into a task-tuned retrieval tool.

\paragraph{Disclosure of AI Usage}
This paper was assisted by Claude; 
however, any assistance provided by AI tools was limited to code generation and language editing under the author's supervision.
The authors have verified the correctness and integrity to the best of their knowledge.

\bibliographystyle{ACM-Reference-Format}
\bibliography{mybibliography}
\end{document}